\documentstyle[onecolumn,psfig,rotating]{mnv2}

\def\lsim{~\rlap{$<$}{\lower 1.0ex\hbox{$\sim$}}}
\def\bsim{~\rlap{$>$}{\lower 1.0ex\hbox{$\sim$}}}

\def\dd{{\rm d}}
\def\ln{{\rm ln}}

\def\pmb#1{\setbox0=\hbox{#1}%
\kern-.025em\copy0\kern-\wd0
\kern.05em\copy0\kern-\wd0
\kern-.025em\raise.0433em\box0}

\def\db {\delta_b}
\def\dx {\delta_x}
\def\ldb {\Delta_b}
\def\ldx {\Delta_x}
\def\trh {\frac{3}{2}}
\def\sterm {s^2+\frac{1}{2}s}
\def\omtot {f_x \ldx + f_b \ldb}
\def\rp {{\tau}}

\begin{document}
\title[Analytic solutions for  coupled perturbations]
{Analytic solutions for coupled linear  perturbations}
\author[Nusser ]{Adi Nusser
\\
Physics Department,
Technion, Haifa 32000, Israel\\
E-mail: adi@physics.technion.ac.il
}
\maketitle

\begin{abstract}
Analytic solutions for the evolution of cosmological linear density
perturbations in the baryonic gas and collisionless dark matter are
derived. The solutions are expressed in a closed form in terms of
elementary functions, for arbitrary baryonic mass fraction.  They are
obtained assuming $\Omega=1$ and a time independent comoving Jeans
wavenumber, $k_J$.  By working with a time variable $\tau\equiv
\ln(t^{2/3})$, the evolution of the perturbations is described by
linear differential equations with constant coefficients. The new
equations are then solved by means of Laplace transformation assuming
that the gas and dark matter trace the same density field before a sudden
heating epoch.  In a dark matter dominated Universe, the ratio of
baryonic to dark matter density perturbation decays with time roughly like
$\exp(-5\tau/4)\propto t^{-5/6}$ to the limiting value
$1/[1+(k/k_J)^2]$.  For wavenumbers $k>k_J/\sqrt{24}$, the decay is
accompanied with oscillations of a period $ 8\pi/\sqrt{24 (k/k_J)^2 -1}$ in
$\tau$.  In comparison, as $\tau $ increases in a baryonic matter
dominated Universe, the ratio approaches $1-(k/k_J)^2$ for $k\le k_J$,
and zero otherwise.
\end{abstract}
\section {Introduction}

Several methods (e.g., Croft et al. 1998; Gnedin 1998; Nusser \&
Haehnelt 1999) have recently been proposed for extracting
information on the mass density field from the Lyman-$\alpha$ forest.
The underlying physical picture behind these method is that the
absorbing Neutral hydrogen in the low density intergalactic medium
(IGM) is tightly related to the mass density field on scales larger
than the Jeans length. Below the Jeans length gas pressure segregates
the baryons from the total mass fluctuations. On scales near the Jeans
length, the evolution of the baryonic perturbation can affect
estimates of the clustering amplitude from observations of the
Lyman-$\alpha$ forest.

Hydrodynamical simulations (Petitjean, M\"ucket \& Kates 1995; Zhang,
Anninos \& Norman 1995; Hernquist et al.~1996; Miralda-Escud\'e et
al.~1996, Theuns et al. 1998) and semi-analytic models (e.g., Bi
et. al. 1992, Gnedin \& Hui 1996) of the IGM have been successful at
explaining observations of the forest.  Despite the success of the
simulations, it is usually difficult to use them to study in detail
the evolution of the gas below Jeans length (Theuns et al. 1998).  The
equations governing the evolution of baryons and dark matter in the
nonlinear regime are extremely difficult to solve, even for special
configurations like spherical collapse.  Fortunately, since most of
the IGM is of moderate density, linear analysis can be a suitable tool
for understanding the evolution of the baryons (Gnedin \& Hui 1998).
Here we derive analytic solutions to the linear equations in a flat
universe without a cosmological constant.  Although the linear
equations can readily be numerically integrated under a variety of
conditions (Gnedin \& Hui 1998), analytic treatment offers better
understanding of the equations.  Further, the paucity
of analytic solutions makes their  pursue worthwhile, even if tedious at
times.
The analytic solutions we derive here are  
subject to the condition that the  baryonic and dark matter 
trace the same density and
velocity fields before a sudden reionization epoch.
After reionization the temperature of the IGM is
assumed to be inversely proportional to the scale factor so that the
comoving Jeans length is constant.

The paper is organized as follows. 
In section 2 we cast the equations in the form 
of linear differential equations with constant coefficients.
In Section 3 we present the solutions to these equations for 
several cases. We conclude in Section 4. 

\section{The linear equations}
Let $\dx(t,k)$ and $\db(t,k)$ be , respectively, the Fourier modes
of   baryonic and dark matter 
density fluctuations. Let also $f_X$ and $f_b=1-f_x$ be the 
the mean mass fractions of  these two types of matter. 
We will restrict the analysis to perturbations in a flat universe 
without a cosmological constant.
The linear equations governing the evolution 
of $\db$ and $\dx$ are (e.g., Bi et. al. 1992, Padmanabhan 1993, 
Gnedin \& Hui 1998),
\begin{eqnarray}
\frac{\dd^2 \dx}{\dd t^2} +2H\frac{\dd \dx}{\dd t}&=& 
\frac{3}{2}H^2\left(f_x \dx + f_b \db \right) \cr
\frac{\dd^2 \db}{\dd t^2} +{2}H\frac{\dd \db}{\dd t}&=& 
\frac{3}{2}H^2\left(f_x \dx + f_b \db \right) - \frac{3}{2}H^2 
\left(\frac{k}{k_J}\right)^2  \db \; ,
\label{lint}
\end{eqnarray}
where $a(t)\propto t^{2/3}$ is the scale factor, $H(t)=2/(3t)$ the
Hubble function, and $k_J$ is the comoving Jeans wavenumber related to the
speed of sound $c_s$ and the mean total (baryons plus dark) density,
$\bar \rho=3H^2/(8\pi G)$, by
\begin{equation}
k_J=\frac{a}{c_s}\sqrt{4\pi G \bar \rho}=\sqrt{\frac{3}{2}}\frac{a H}{c_s} .
\end{equation}

For $f_x$ close to unity the gravity of the baryons is negligible and
$\dx \propto a$. If also $k_J=const$ (i.e., $c_s^2\propto 1/a$) and we
impose, at some initial time $t_i$, the condition
$\db(t_i)=\dx(t_i)/(1+(k/k_J)^2)$, then $\db(t) =\dx(t)/(1+(k/k_J)^2)$
at any later time, $t$.  In the next section we will show how to solve
these equations with $0<le f_x\le $ assuming that 
$\db(t)=\dx(t)$ for $t\le t_i$, which is appropriate for a
sudden reionization of the IGM at $t_i$.

In order to solve these equations we work with a new time 
variable $\rp \equiv \ln (a)$ instead of $t$ (Nusser \& Colberg 1998).
In terms of $\tau$ the equations  become,
\begin{eqnarray}
\frac{\dd^2 \dx}{\dd \rp^2} +\frac{1}{2}\frac{\dd \dx}{\dd \rp}&=& \frac{3}{2}\left(f_x \dx +
f_b \db \right) \cr 
\frac{\dd^2 \db}{\dd \rp^2} +\frac{1}{2}\frac{\dd \db}{\dd \rp}&=& \frac{3}{2}\left(f_x \dx +
f_b \db \right) -\frac{3}{2} \kappa^2 \db \; ,
\label{lintau}
\end{eqnarray}
where have defined  $\kappa=k/k_J$.

When $\kappa$ is constant, the differential equations (\ref{lintau}) 
are linear  with constant coefficients and they can be solved
by means of Laplace transformation.
Since Laplace transforms are seldom used in cosmological studies, it seems 
prudent to briefly review
their basic properties which are relevant to us.  
We refer the reader to Arfkin (1985) and references therein
for mathematical details.
The Laplace transform, $f(s)$, of a function $F(t)$, where $t\ge 0$,
is defined as 
\begin{equation}
f(s)\equiv {\cal L}\{F(t)\}=\int_0^\infty \exp\left(-st\right) F(t) \dd t .
\label{laplace:def}
\end{equation} 
We will need the Laplace transforms of first and second derivatives of a
function.  Using (\ref{laplace:def}) these transforms can be related to the
$f(s)$ by
\begin{eqnarray}
{\cal L} \{ F'(t)\} &=& s f(s)- F(0) \cr
{\cal L} \{ F''(t)\} &=& s^2 f(s)- s F(0) - F'(0) \; ,
\label{laplace:derv}
\end{eqnarray}
where the prime and double prime denote first and second order 
derivatives, respectively. 
The Bromwich integral expresses  $F(t)$ in terms of $f(s)$
as 
\begin{equation}
F(t)=\frac{1}{2\pi i}\int_{\gamma -i\infty}^{\gamma+i\infty}
\exp\left(s t\right) f(s)\dd s 
\label{bowich}
\end{equation}
where $i=\sqrt{-1}$ and $\gamma$ is a real number chosen so that
all  poles of $f(s)$ lie, in the complex plane, to the left of the 
vertical line defining the integration path. Therefore, by the residue
theorem we have
\begin{equation}
F(t)=\sum \left[
{\rm residues} \; {\rm of} \; \exp\left(s t\right)f(s)
\right]
\label{resid}
\end{equation}
As an example consider $f(s)=1/(s-s_1)(s-s_2)$ which has two simple poles at 
$s=s_1 $ and $s_2$. The residues of $\exp({s t}) f(s)$ at these poles are 
$\exp({s_1 t}) /(s_1-s_2)$ and $-\exp({s_2 t}) /(s_1-s_2)$ so that, by (\ref{resid}),
$F(t)=[\exp({s_1 t})- \exp({s_2 t})]/(s_1-s_2)$. If $s_1=s_2$,  the function 
has a pole of order two at $s_1$. The residue 
in this case is $\dd[(s-s_1)^2 \exp({st})f(s))]/\dd s$ evaluated at $s=s_1$. 
Therefore,  $F(t)=t \exp({s_1 t})$.

\section{The solutions}

Denote by $\ldx$ and $\ldx$ the Laplace transforms of $\dx$ and $\db$,
respectively. By taking the Laplace transform of (\ref{lintau}) we can
obtain relations between $\ldx$ and $\ldb$.  The initial conditions
are contained in the Laplace transforms of the first and second
derivatives of the densities.  So first we have to specify in
mathematical terms our choice for the initial conditions.  For
simplicity of notation we fix the initial conditions at $\tau=1$
assuming that before that time the temperature of the baryonic fluid
is zero, i.e., $\kappa=0$. The initial conditions are fixed by the
values of $\dx$ and $\db$ and their first derivatives at $\tau=1$.
Before $\tau=1$, we have $\dx=\db=\exp(\tau)$, ignoring the decaying
mode and setting arbitrarily $\dx(\tau=1)=1$.  The first derivatives
of $\dx$ and $\db$ are therefore equal to unity at $\tau=1$. This
fixes the initial conditions necessary for solving
(\ref{lintau}). Although we will present solutions satisfying only
these initial conditions, we will, for completeness, write the Laplace
transformation of (\ref{lintau}) for $\db=\alpha\dx $ and $\dd \db
/\dd \tau=\alpha \dd \dx /\dd \tau$, at $\tau=1$.  
Then the Laplace transformation of (\ref{lintau}) yields
\begin{eqnarray}
\left( \sterm \right) \ldx &=& 
\trh \left( \omtot \right) +s +\trh\cr 
\left( \sterm \right) \ldb &=& 
\trh \left( \omtot \right) -
\trh \kappa^2 \ldb + \alpha s +\trh \alpha \; ,
\label{laplace:lin}
\end{eqnarray}
where have have  used (\ref{laplace:derv})  to computed the 
transforms of the first and second derivatives of $\dx$ and $\db$.
For $f_x=1$ the first of these equations yields
\begin{equation}
\ldx=\frac{1}{s-1}\; ,
\label{ldxfx}
\end{equation}
which is the Laplace transform of $\exp(\tau)$. 
If we take  $\alpha=(1+\kappa^2)^{-1}$ and 
substitute (\ref{ldxfx})  in 
the second equation of (\ref{laplace:lin}) we get 
\begin{equation}
\ldb=\frac{1}{s-1}\frac{1}{1+\kappa^2}=\frac{\ldx}{1+\kappa^2} \; ,
\end{equation}
which leads to the well known solution $\db=\dx/(1+\kappa^2)$. 

Subsequently we will present solutions only  for $\alpha=1$.
In this case,
equations (\ref{laplace:lin}) yield %IMPORT
\begin{equation}
\ldx= \frac{\left( s +\trh\right) \left(\trh \kappa^2+\sterm\right)}
{\left(\sterm\right)
\left(
\trh \kappa^2+\sterm-\trh
\right) -\frac{9}{4}f_x\kappa^2} \; ,
\label{dxg} 
\end{equation}
and
\begin{equation}
\ldb=\frac{\sterm}{\trh \kappa^2 
+\sterm}\ldx  \; .
\label{dbdx}
\end{equation}

Before solving these equations for any value
of $f_x $ in the range 0--1,  it is instructive to 
examine the solutions for the special values $f_x=1$ and $0$.

\subsection{ Case I: $f_x=1$}
In this case $\ldx=(s-1)^{-1}$ and  equation 
(\ref{dbdx}) can be written in the form 
\begin{equation}
\ldb=\frac{\sterm}{(s-1)(s-s_{_-})(s-s_{_+})}
\end{equation}
where  $s_{_\pm}$ are the roots of $3\kappa^2/2+s^2+s/2$. They
are given by
\begin{equation}
 s_{_\pm}=-\frac{1}{4}\left(1 \pm \chi\right) 
\qquad ; \qquad \chi^2={1 -24 \kappa^2}
\end{equation}
We will deal with the case $ \chi^2=0$ at the end of this subsection.
For $\chi^2 \ne 0$, 
all three poles of $\ldb$
are simple and so,
by (\ref{resid}), its inverse transform is
\begin{equation}
\db=\frac{\exp\left(\rp\right)}{1+\kappa^2} +
\frac{1}{s_{_-}-s_{_+}}\frac{\kappa^2}{1+\kappa^2} 
\biggl[
\left(s_{_-}-1\right) \exp\left(s_{_+} \rp\right)  - \left(s_{_+} -1\right)
\exp\left(s_{_-} \rp\right)
\label{dbcon}
\biggr]
\end{equation}
For $\chi^2>0$ the roots $s_{_\pm}$ are real and the 
result is 
\begin{equation}
\db=\frac{\exp\left(\rp\right)}{1+\kappa^2} +
\frac{1}{2\chi}\frac{\kappa^2}{1+\kappa^2}
\biggl[\left(\chi-5\right) \exp\left(-\frac{\chi}{4}\rp\right) 
+\left(5+\chi\right)\exp\left(+\frac{\chi}{4}\rp\right)\biggr]
\exp\left(-\frac{\rp}{4}\right)
\label{dbexp}
\end{equation}
The first term is the solution given in the previous section 
for $\alpha=1/(1+\kappa^2)$. 
The maximum value  $\chi^2 $ attains is unity, so
the second and third
terms are always decaying. 
If  $\chi^2 <0$, then (\ref{dbcon}) gives the  solution 
\begin{equation}
\db=\frac{\exp\left(\rp\right)}{1+\kappa^2} +
\frac{1}{\tilde \chi}\frac{\kappa^2}{1+\kappa^2}
\biggl[5\sin\left(\frac{\tilde \chi}{4}\rp\right) +
\chi\cos\left(\frac{\tilde \chi}{4}\rp\right)
\biggr]\exp\left(-\frac{\rp}{4}\right)
\label{dbsin}
\end{equation}
where $\tilde \chi$ is the imaginary part of $\chi$.  The solution
shows an oscillatory behavior with a period of $16\pi/\tilde \chi$. The
envelope of these oscillations decays like $\exp(-\tau/4)\propto
t^{-1/6}$.

We deal now with the case $\chi^2=0$, which occurs for $\kappa^2=1/24$.
Here special care is needed because $s_{_-}=s_{_+}=-1/4$.  However 
the contribution of the pole at $s=-1$ to the Bromwich integral 
remains unchanged and the
contribution of the  second order pole at $s=-1/4$ is simply 
the first derivative of $\exp({s\tau})
(s+1/4)^2\ldb$ at $s=-1/4$. The result is
\begin{equation}
\db=\frac{24}{25}\exp\left(\tau\right) +\frac{1}{20}\left(\tau + \frac{4}{5}\right)
\exp\left(-\frac{\tau}{4}\right) \; .
\end{equation}
The first term on the left is the familiar $\dx /(1+\kappa^2)$ evaluated 
at $\kappa^2=1/24$.
The expression can also be derived by taking the limit $\chi^2\rightarrow 0$ 
in either (\ref{dbexp})   or (\ref{dbsin}).

In the limit  $\tau\rightarrow \infty$ 
the ratio $\db/\dx$ is $1/(1+\kappa^2)$.
% for $\kappa \le 1$ and zero otherwise.

\subsection{ Case II: $f_x=0$}
This is equivalent to ignoring the gravity of the 
dark matter. Of course here only the behavior of the 
perturbation in the baryons is relevant since  the dark matter 
plays no role. 
However,  for the sake of completeness  and comparison with other situations
we will solve for the dark matter fluctuations as well.

We first find the solution for $\dx$.
If  $f_x=0$, we can express (\ref{dxg}) in 
terms of $s_{_\pm}$, the roots of
$3\kappa^2/2+s^2+s/2-3/2$, as
\begin{equation}
\ldx= \frac{\left( s +\trh\right) \left(\trh \kappa^2+\sterm\right)}
{s\left(s+\frac{1}{2}\right)
\left(s-s_{_-}
\right)
\left(s-s_{_+}
\right)
} \; ,
\end{equation}
where 
\begin{equation}
 s_{_\pm}=-\frac{1}{4}\left(1 \pm \chi\right) 
\qquad ; \qquad \chi^2={25 -24 \kappa^2}
\end{equation}

If  $\kappa\ne 1$, then the function $\ldb$ has three simple 
poles at $s=0$, $s_{_-}$, and $s_{_+}$. 
So for $\kappa \ne 1$ and $\chi^2>0$, Bromwich integral 
yields
\begin{equation}
\dx=
\frac{\kappa^2}{1-\kappa^2}\biggl[
2 \exp\left(-\frac{\tau}{2}\right)-3
\biggr]
+\frac{1}{2\chi}\frac{1}{1-\kappa^2}
\biggl[\left(\chi-5\right)\exp\left(-\frac{\chi}{4}\tau\right)
+\left(\chi+5\right)\exp\left(+\frac{\chi}{4}\tau\right)\biggr]
\exp\left(-\frac{\tau}{4}\right) \; .
\label{dxs2}
\end{equation}
The expression for $\db$ is
\begin{equation}
\db=\frac{1}{2\chi}\biggl[
\left(\chi -5\right) \exp\left(-\frac{\chi}{4}\tau\right)+
\left(\chi+5\right) \exp\left(\frac{\chi}{4}\tau\right)\biggr]
\exp\left(-\frac{\tau}{4}\right) \; .
\label{dbs2}
\end{equation}

When $\kappa=1$ the solution can be found either by taking the limit
$\kappa\rightarrow 1$ in the previous two 
expressions or by direct evaluation of the Bromwich integral
with a two poles of order two  at $s=0$ and $-1/2$. The result is   
\begin{equation}
\dx=-27+28\exp\left(-\frac{\tau}{2}\right)+
3\tau\left[3+2 \exp\left(-\frac{\tau}{2}\right) \right] \qquad {\rm and} \qquad
\db=3-2 \exp\left(-\frac{\tau}{2}\right)  \; .
\end{equation}
This implies that $\dx$ grows
linearly with $\tau$ at late times
while $\db$ reaches an asymptotic value of $3$.

The oscillatory behavior of $\db$ and $\dx$ appears when  $\chi^2<0$,
i.e., for $\kappa^2 >25/24$. The expressions in this case 
can be obtained by replacing $\chi$ with $i\tilde \chi$ in 
(\ref{dxs2}) and (\ref{dbs2}). For $\kappa^2 >\!> 25/24$, the solution
coincides with that given in Padmanabhan (1993)

In the limit  $\tau\rightarrow \infty$ 
the ratio $\db/\dx$ is $1-\kappa^2$
for $\kappa \le 1$ and, zero otherwise.

\subsection{ Case III: $0<f_x<1$}
Again we first derive $\dx$.
The denominator and numerator in (\ref{dxg}) do not 
have any common roots. 
Then the poles of $\ldx$ are the roots of the denominator.
We find these roots as follows. Denote $y(s)=s^2+s/2-3/2$ and equate
denominator to zero to obtain
\begin{equation}
y^2+\trh y \left(1+\kappa^2\right) +\frac{9}{4}f_b \kappa^2=0
\end{equation}
where we have used $1-f_x=f_b$.
This equation is   satisfied for the following values of $y$,
\begin{equation}
y_{p,m}=-\frac{3}{4}\left(1+\kappa^2\right)\left(1\pm \Xi\right) 
\qquad ; \qquad \Xi^2=1-4f_b\frac{\kappa^2}{\left(1+\kappa^2\right)^2} \; ,
\end{equation}
where the subscripts $p $ and $m$ correspond to the plus and minus sign,
respectively.
So the roots of the denominator in (\ref{dxg}) are 
the values of $s$ which make $y(s)=y_{p,m}$.
Let  $s_{p,\pm}$ and $s_{_{m,\pm}}$ be the roots of  
$y(s)-y_{p}=0$ and $y(s)-y_{m}=0$, respectively. They are given by
\begin{eqnarray}
s_{_{m,\pm}}=-\frac{1}{4}\left(1\pm\chi_m\right) &;& 
{\chi_m}^2=25-12 
\left(1+\kappa^2\right)\left(1+\Xi\right) \; ,\cr
s_{_{p,\pm}}=-\frac{1}{4}\left(1\pm\chi_p\right) &;& 
{\chi_p}^2=25-12
\left(1+\kappa^2\right)\left(1-\Xi\right) \; .
\end{eqnarray}

Excluding the values $f_b=0$ and $1$, which we have considered in the
previous subsections, we have $0<\Xi^2<1-f_b$. This ensures that all
four roots, $s_{_{m,\pm}}$ and $s_{_{p,\pm}} $, are distinct. Also,
$1<\chi_p^2<25$ for any $\kappa$, so the roots $s_{_{p,\pm}}$ are
real. On the other hand, $0<\chi^2_m<1$ when $\kappa^2
<25/(600-576f_b)$, and negative otherwise. So $s_{_{m,\pm}}$ can be
complex.

For $\chi_m^2>0$, the Bromwich integral yields
\begin{equation}
\dx=\frac{24\kappa^2+\chi_p^2-1}
{2\chi_p
\left(
\chi_p^2-\chi_m^2
\right)}
\biggl[\left( \chi_p-5\right)\exp\left(-\frac{\chi_p}{4}\rp \right)
+\left( \chi_p+5\right)\exp\left(+\frac{\chi_p}{4}\rp \right)\biggr]
\exp\left(-\frac{\tau}{4}\right)
 + (\chi_p \longleftrightarrow \chi_m) \; ,
\end{equation}
with the second term on the r.h.s is obtained 
interchanging $\chi_p$  and $\chi_m$ in the first term.
This result can be extended  to $\chi_m^2<0 $ 
by writing $\chi_m=i \tilde \chi_m$ where $\tilde \chi_m$ is real.

Using (\ref{dbdx}) we similarly obtain  
$\db$
\begin{equation}
\db=\frac{\chi_p^2-1}
{2\chi_p
\left(
\chi_p^2-\chi_m^2
\right)}
\biggl[\left( \chi_p-5\right)\exp\left(-\frac{\chi_p}{4}\rp \right)
+\left( \chi_p+5\right)\exp\left(+\frac{\chi_p}{4}\rp \right)\biggr]
\exp\left(-\frac{\tau}{4}\right)
 +  (\chi_p \longleftrightarrow \chi_m) \; .
\end{equation}

To visualize  these solutions,
we show in Fig.1 the density evolution for various values of $f_x$ and
$\kappa$. 
The dark matter curve for $f_x=0.9$ is very close to 
$\exp(\tau)$. For $\kappa=5$ the baryonic perturbations
for both values of $f_x$ show oscillations with similar period and amplitude.
This is simply because for high $\kappa$, the evolution is mainly dictated 
by pressure forces.

In Fig.2 we the solid lines represent the ratio
$\db(\kappa)/\dx(\kappa)$ as a function of $\kappa$ at different
$\tau$ designated in the plot by the redshift, $z$.  The initial
conditions are satisfied at $z=6$ and $f_x=0.9$ was taken.
Also plotted, as the dotted line in each panel, the function
$1/(1+\kappa^2)$ which represents the limiting solution as $\tau\rightarrow \infty$.  The analytic
curves show more oscillations as they get closer to the limiting
solution, when the redshift is decreased.

In many applications (e.g., Bi et al 1992, Bi \& Davidsen 1997, Nusser
\& Haehnelt 2000) the limiting ratio $1/(1+\kappa^2)$ is often used to
filter the mass power spectrum in order to generate density
fluctuations in the gas. As pointed out by Gnedin \& Hui (1998) this
may lead to a significant bias in statistics of the gas density.  As
an illustration of this bias we compute the rms values of the gas
density by filtering a scale free mass power spectrum of slope
$n=-2.5$ with the ratio $(\db/\dx)^2 $ given from the analytic, and
the limiting solutions, respectively.  Fig.3 shows the ratio of the
former to the latter rms value.  As in the previous figure, $f_x=0.9$
and initial conditions satisfied at $z=6$. Using the filter
$1/(1+\kappa^2)$ can seriously underestimate the amplitude of gas
density fluctuations.  Only when we approach $z=0$ the ratio gets
close to to unity.
 
\section{Summary}
We have found analytic solutions to the linear equations governing the
evolution of baryonic and dark matter under four assumptions. First,
the Universe is flat without cosmological constant.  Second,
sudden reionization of the IGM. Third, the temperature of
the low density IGM drops like $1/a$  so that the comoving Jeans 
length is time independent. Fourth, before reionization the IGM is cold and 
the baryonic and dark matter trace the same density and velocity fields.

Of these assumptions, only the fourth has a physical basis, at least
before any heating has occurred and when the IGM temperature is low.
This is also the only assumption which if changed, the equations can
still be readily solved by Laplace transformation.  Unfortunately,
relaxing any of the other assumptions complicates the analytic
treatment of equations (\ref{lintau}) by means of Laplace
transformation.  For example, suppose that the Jeans wavenumber changes
with time according to $a^\beta$. Then the Laplace transformation of
the term involving $k_J$ will yield $\ldb$ at $(s+\beta)$ while other
terms involve $\ldb(s)$.

Yet the solutions can be useful for semi-analytic modeling of the
IGM. They offer a convenient improvement over the commonly used filter
$1/[1+(k/k_J)^2]$ for generating gas fluctuations associated with a
given mass density field.

The analytic solutions presented here were verified by a comparison
with the solutions obtained by numerical integration of equations
(\ref{lint}).  All numerical and analytic solutions agreed up to the
numerical accuracy.

\section{acknowledgement}
This research was supported by a grant from the Israeli Science Foundation.

\begin{figure}
\centering
\begin{sideways}
\mbox{\psfig{figure=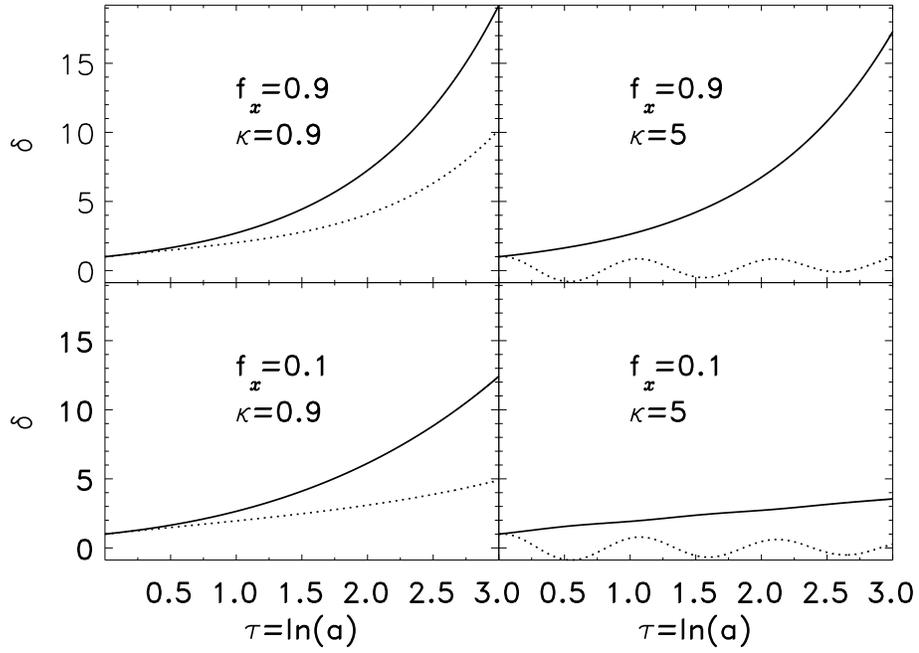,height=6.0in}}
\end{sideways}
\caption{ Curves of $\dx(\tau)$ (solid lines) and $\db(\tau)$ (dotted) 
for various values of $f_x$ and $\kappa=k/k_J$. }
\label{fig:fig1}
\end{figure}                     

\begin{figure}
\centering
\begin{sideways}
\mbox{\psfig{figure=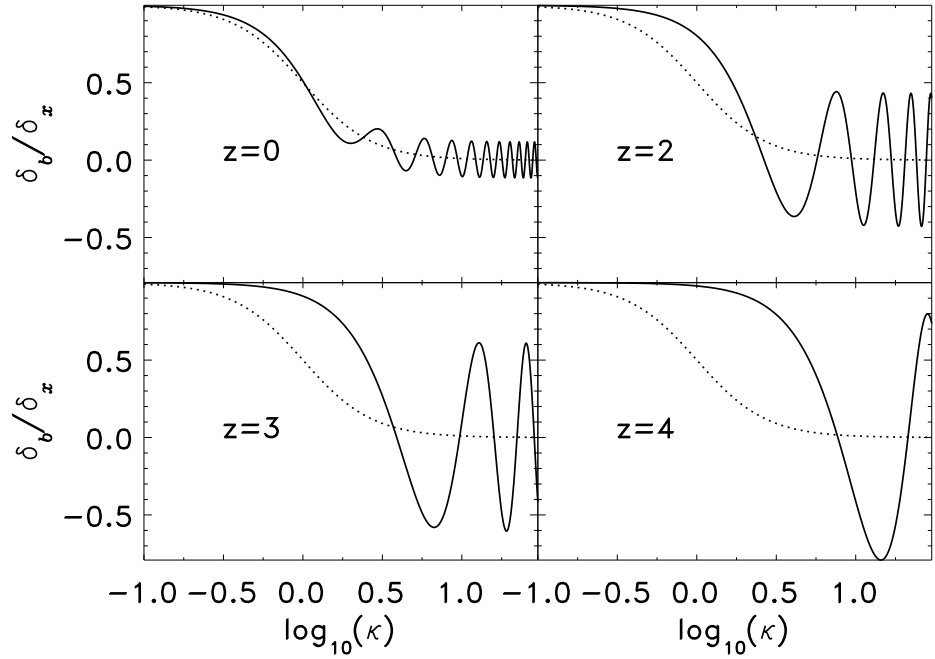,height=6.0in}}
\end{sideways}
\caption{ 
Ratios of baryonic to dark matter density as a 
function of $\kappa$  at different times, indicates by the redshift, $z$,
in each panel.
The solid lines are  the analytic solutions 
obtained with $f_x=0.9$ and where the initial conditions are satisfied 
at $z=6$.
The dotted line in each panel shows  
$1/(1+\kappa^2)$, the ratio corresponding to the limiting solution.}
\label{fig:fig2}
\end{figure}                     

\begin{figure}
\centering
\begin{sideways}
\mbox{\psfig{figure=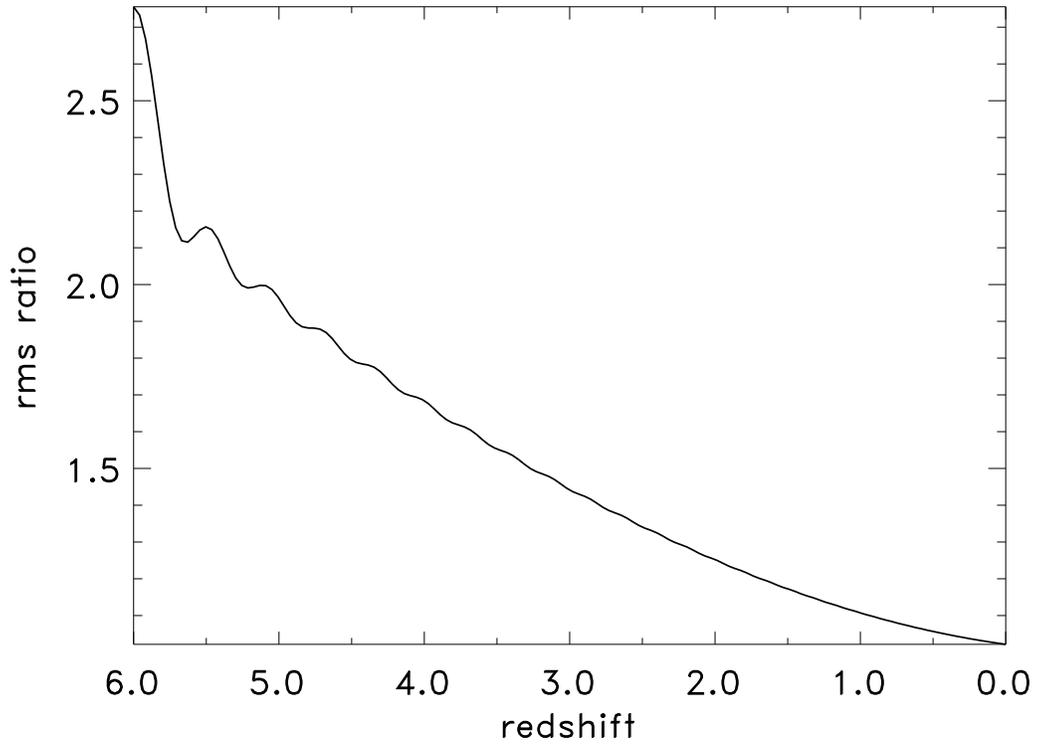,height=6.0in}}
\end{sideways}
\caption{ The ratio of the baryonic density $rms$ values 
computed by filtering
the mass power spectrum with $(\db(\tau,\kappa)/\dx(t,\kappa))^2$ in the
analytic solution to that obtained by filtering with
$1/(1+\kappa^2)^2$.
A scale free mass power spectrum with slope $n=-2.5$ is assumed.}
\label{fig:fig3}
\end{figure}                     

\end{document}